\documentclass[aps,twocolumn,nofootinbib,preprintnumbers,pra,10pt]{revtex4}
\usepackage{amsfonts,amsthm,amsmath,amssymb,graphicx,hyperref}
\usepackage{braket,url,color}

\newcommand{\be}{\begin{equation}}
\newcommand{\bea}{\begin{eqnarray}}
\newcommand{\ee}{\end{equation}}
\newcommand{\eea}{\end{eqnarray}}

\newcommand{\bes}{\begin{equation*}}
\newcommand{\ees}{\end{equation*}}
\newcommand{\beas}{\begin{eqnarray*}}
\newcommand{\eeas}{\end{eqnarray*}}

\newcommand{\bmat}{\begin{bmatrix}}
\newcommand{\emat}{\end{bmatrix}}

\begin{document}

\title{Renormalized Circuit Complexity}

\author{Arpan Bhattacharyya$^{1,2}$\email{abhattacharyya@iitgn.ac.in}, Pratik Nandy$^{3}$\email{pratiknandy94@gmail.com} and Aninda Sinha$^{3}$\email{asinha@iisc.ac.in}}
\affiliation{
${^1}$Indian Institute of Technology, Gandhinagar, Gujarat 382355, India\\
${^2}$Center for Gravitational Physics,  Yukawa Institute for Theoretical Physics, Kyoto University, Japan
  \\
 ${^3}$Centre for High Energy Physics, Indian Institute of Science, C.V. Raman Avenue, Bangalore, India. }

\begin{abstract}
We propose a modification to Nielsen's circuit complexity for Hamiltonian simulation using the Suzuki-Trotter (ST) method,  which provides a network like structure for the quantum circuit. This leads to an optimized gate counting linear in the geodesic distance and spatial volume, unlike in the original proposal. 
The optimized ST iteration order is correlated with the error tolerance and plays the role of an anti-de Sitter (AdS) radial coordinate.  The density of gates is shown to be monotonic with the tolerance and a holographic interpretation using path-integral optimization is given. 
\end{abstract}

\maketitle

{\it \noindent Introduction---} One of the key questions in quantum computing is to find efficient quantum circuits which can simulate Hamiltonian evolution.  Nielsen and collaborators showed that this problem can be geometrized in such a way that the  minimum number of quantum gates is related to the geodesic length between the identity operator $I$ and the desired unitary $U$ in the ``circuit space'' \cite{Niel1,  Niel2,Niel3,Niel4}. 

In \cite{Niel1}, an  explicit procedure was given to construct the circuit. The first step is to define a control Hamiltonian $\tilde H(s)$ and split it into an ``easy" part and a ``hard" part where the latter involves gates difficult to make in a laboratory. Here $s$ parametrizes the circuit depth. Then one writes down a cost function which is minimized to obtain a geodesic in circuit space which tells us how the gates should be arranged in an optimum manner. The hard gates are penalized using penalty factors (which we will generically denote by $p$) thereby increasing the cost in that direction.  The geodesic length is denoted by $d(I,U)$ and in general depends on $p$. A specific cost functional, that is frequently used, induces a  Riemannian metric on the circuit space \cite{JM,JM2a,JM10, JM1,JM2,JM3,JM4,JM5,JM6,JM7,JM8,JM8a,JM9a,JM9,JM11,JM12,JM13a,JM14, JM14a,JM15,JM16}. In \cite{sussrec}, this geometry was called the ``complexity geometry''.  In recent literature, this has played a crucial role to compare with holography \cite{Susskind4, Susskind5, Susskind1,Hol5,Carmi:2017jqz,Hol6,Susskind1a,Susskind2,Susskind3,Susskind6,Susskindlec,Alishahiha:2015rta,Hol1,Hol2,Hol3,Hol7,Hol12,Kim, Hol18, Hol18a,Hol21,Hol24,Flory,Flory1,Flory2,Hol26,Hol27,Hol28,rub}. {\it However, the total number of gates in \cite{Niel1} is not just given by $d(I,U)$; in fact in \cite{Niel1}, it is not even linear in $d(I,U)$ as we will review below.}

 To count the total number of gates, Nielsen \cite{Niel1} first constructs a projected Hamiltonian $\tilde H_{p}(s)$ by simply deleting the hard gates from the control Hamiltonian evaluated on the geodesic solution. The corresponding projected unitary $\tilde{U}_p$ provides a good approximation to the target $U$ up to some error. The next step according to \cite{Niel1} is to divide the total path $d(I,U)\equiv d$ into $N = d/\Delta$ steps of equal interval $\Delta$, and for each of these intervals, we define an averaged Hamiltonian, $\bar H = \frac{1}{\Delta} \int_0^{\Delta} ds\, \tilde{H}_P(s)$ with the average unitary $\bar U =e^{- i\,\bar H \Delta }$ (which is eventually applied $N$ times). This step bypasses the need to work with path-ordered expressions. The final step is to further divide the interval $[0, \Delta]$ into $r = 1/\Delta$ intervals with each of length $\Delta^2$ and approximate the average unitary by quantum gates using the Lie-Trotter formula \cite{nielsenchuang}. Putting all these results together and assuming all penalty factors to be identical (without loss of generality), one obtains \cite{Niel1}
the total number of gates required to synthesize the unitary as $N_{\mathrm{gates}} = O(m^3 d^3/\delta^2)$ \cite{Niel1} where $m$ is the number of easy terms in the Hamiltonian and $\delta$ is the specified tolerance. If the Hamiltonian is ``geometrically local" \cite{presslec2}---$g$-local in short---which means that all the qubits can be arranged inside a fixed $D$-dimensional volume $V$, then it can be shown following \cite{presslec2}, that
 $N^{\mathrm{local}}_{\mathrm{gates}} = O(m^2 d^3/\delta^2)$, so the dependence on $m$ is $m^2$, not $m^3$. Now, since $m\propto V$, we have $N^{\mathrm{local}}_{\mathrm{gates}} = O(V^2 d^3/\delta^2)$. The dependence of $V$ as found in \cite{Niel1} is thus unlike holographic proposals, which have suggested that complexity should be just linear in $V$ \cite{Susskind4,Susskind5}. Clearly one should be able to do better since effective field theory reasonings, that work so well to describe nature, suggest that the scaling should be linear in the spatial volume.
 We will give an improvement below which will make the optimized number of gates linear in $V$
 -- moreover, as we will see, this improvement seems to tie up with holography in an interesting way. 
 
\begin{figure}[h]
\centering
\includegraphics[scale=0.3]{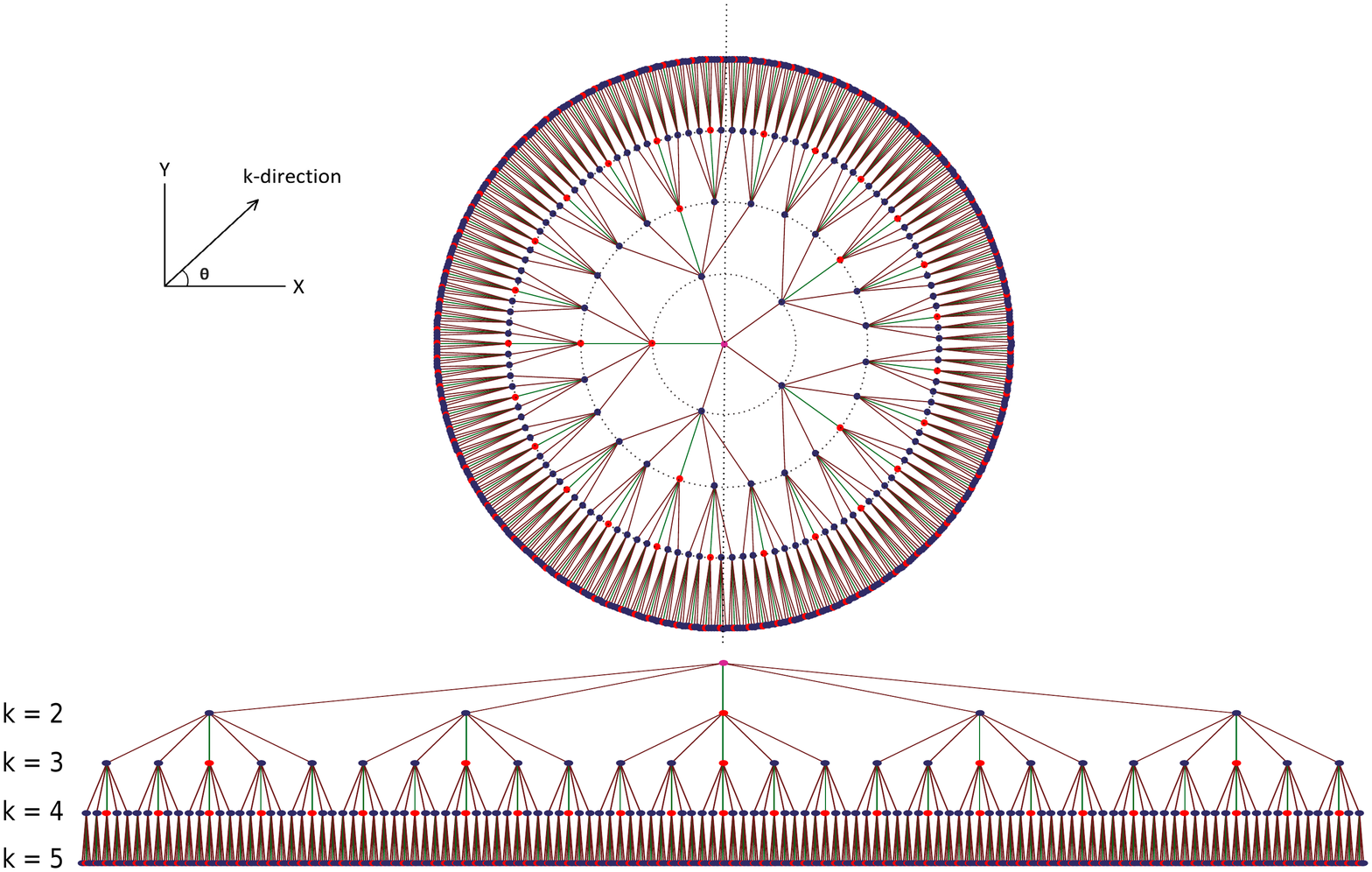}
\caption{The ST ``holographic'' network. The circuit above is a ``compactified'' version of the circuit below and is a pictorial representation of the ST recursion relation.} \label{ckt1}
\end{figure}

{\it \noindent Improvement---} Now the final step used above admits an immediate improvement. Instead of the Lie-Trotter formula used in \cite{Niel1}, we can use the $k$-th order integrators of the Suzuki-Trotter (ST) method \cite{Suzuki1,Suzuki2} to approximate the circuit constructed by the average unitary. Thus, for any small time interval $\Delta$, the unitary made of the mean Hamiltonian $\bar{H}$ can be approximated by $S_{2k} (-i \Delta/r)$ \cite{Suzuki1, Barry1a, Barry1,Barry2}  which satisfies \cite{foot1}: 
\begin{align}
\begin{split} \bigg|\bigg| e^{- i \sum_{j = 1}^m \bar{H}_j \Delta} - [S_{2 k}(-i \Delta/r)]^r \bigg| \bigg|  \leq \frac{2 \, \kappa_m (2h \, 5^{k-1} \Delta)^{2k+1}}{r^{2k}} \label{st1} 
\end{split}
\end{align}
for $\Delta \to 0$. The factor $\kappa_m$ depends whether we choose $K$-local or $g$-local Hamiltonian \cite{presslec2}. For $K$-local, the number of non-zero commutators $[\bar{H}_a, \bar{H}_b]$ is $O(m^2)$ and in that case $\kappa_m = m^{2k+1}$. However if the Hamiltonian is $g$-local \cite{presslec2}, then we will have $\kappa_m = m$. Here we have also assumed that $\bar{H} = \sum_{j=1}^m \bar{H}_j$, which can be exponentiated easily and can be written in terms of elementary gates and we have $h \geq \max ||\bar{H}_j||$. Here, we have also divided each path interval $[0,\Delta]$ into $r$ intervals and $S_{2k}(-i \Delta/r)$ is given by the recursion relation $S_{2k}(-i \Delta/r) = [S_{2k - 2}(q_k (-i \Delta/r))]^2 \, S_{2k-2}((1-4 q_k)(-i \Delta/r)) \, [S_{2k - 2}(q_k (-i \Delta/r))]^2$ with $q_k = (4 - 4^{\frac{1}{2k-1}})^{-1}$ for $k > 1$ \cite{Barry1a,Barry1} with the initial condition $S_2 (-i \Delta/r) = \prod_{j = 1}^m e^{- i \bar{H}_j \Delta/2 r} \prod_{j' = m}^1 e^{ - i \bar{H}_{j'} \Delta/2 r}$.
The recursion relation naturally gives a network structure of the circuit which can be visualized in the form of the Figure (\ref{ckt1}). The recursion relation involves four $S_{2k-2}$ (solid blue circles in Figure (\ref{ckt1})) with the same argument with another $S_{2k-2}$ with a different argument in the middle (solid red circles in Figure (\ref{ckt1})).  The  magenta solid circle represents the initial $S_2(- i \Delta/r)$. The iteration order $k$ increases in the radial direction and  gives the network depth. As $k$ becomes large, the error is $O(\Delta^{2k+1})$ and becomes small. 
From \eqref{st1}, to have the total error $|| U - U_A || \leq O(\delta)$, where $U_A$ is the simulated unitary, we need $r = \big\lceil 2 h \Delta \, 5^{k - \frac{1}{2}} \big(\frac{4 \, h \, d \, \kappa_m}{5 \,\delta}\big)^{\frac{1}{2k}} \big\rceil,$ where $\lceil ~ \rceil$ is the ceiling function.
Then using this value of $r$ the total number of gates becomes 
\begin{align}
N_{\mathrm{gates}} = O \bigg[h\,  m \, 5^{2k} d^{1+\frac{1}{2k}}\bigg(\frac{4 \, h \, \kappa_m}{5 \, \delta}\bigg)^{\frac{1}{2k}} \bigg] \label{st5a}\,,
\end{align}
which gives a super-linear scaling with $d$ \cite{supp2}. In what follows, we will take the Hamiltonian as $g$-local, hence we take $\kappa_m = m$. Hence the number of gates becomes
\begin{align}
N_{\mathrm{gates}}^{\mathrm{(local)}} = O \bigg[ h \, \Omega^{1+\frac{1}{2k}} 5^{2k} \bigg(\frac{4 \, h }{5 \, \delta}\bigg)^{\frac{1}{2k}} \bigg] \label{st5}\,,
\end{align}
where $\Omega = V d$, and $V \propto m$ is the spatial volume.
 If we wanted to decompose further in terms of a universal set of quantum gates, then the Solovay-Kitaev theorem would give an additional $\ln^{c}(\frac{2 \Omega}{\delta})$  factor with $c \approx 3.97$ \cite{presslec2}. We will drop this factor in what follows. We will also work with the full Hamiltonian rather than the projected one--this will not alter our conclusions.
Notice that for $k\rightarrow \infty$, the dependence of $N_{\mathrm{gates}}$ on $d$ becomes linear. However we can do better!
 
  {\it \noindent Optimization---} Following \cite{Barry1a,Barry1}, one could optimize $k$ in \eqref{st5} to minimize the number of gates, assuming $\Omega$ (i.e., $d$ and $m$) to be independent of $k$--one can think of this assumption as defining a fixed point. Optimization gives

 \begin{align}
N_{\mathrm{gates}}^{\mathrm{opt,(local)}} =O\bigg[  h \, \Omega \exp(4 \ln5 \,k^{\mathrm{opt}})\bigg]  \label{st56}\,,
\end{align}
where  $\Omega=V d$ and \begin{equation}k^{\mathrm{opt}} = \frac{1}{2}\sqrt{\log_5 \bigg(\frac{4 h \Omega }{5 \delta}\bigg)}\,,\label{krel2}\end{equation} 
 From eq. \eqref{st56}, we see that the $\Omega$ dependence now is manifestly linear \cite{com1} for fixed $k^{\mathrm{opt}}$ as suggested by holographic proposals (fixing $k^{\mathrm{opt}}$ is like fixing the AdS cut-off). As the tolerance $\delta\rightarrow 0$, $k^{\mathrm{opt}}\rightarrow\infty$. In other words, the circuit for large $k^{\mathrm{opt}}$ would have lower error and small $k^{\mathrm{opt}}$ would correspond to more coarse-graining. Further for at least small times $t$, it can be shown \cite{supp}, $d\propto t$. 
 
 {\it We suggest that $\Omega \, \exp(4 \ln5 \,k^{\mathrm{opt}} )$ is analogous to the warped volume that one can expect to find in an AdS background! $k^{\mathrm{opt}}$ is the radial cutoff. Changing $k^{\mathrm{opt}}$ corresponds to changing the total number of gates via eq.(4).}
 
 The dependence of $k^{\mathrm{opt}}$  on $\Omega$ is artificial since we can absorb that factor inside $\delta$ and think of $\delta$ as the error tolerance per gate. Once we optimize, it is natural to think in terms of $k^{\mathrm{opt}}$ as the independent variable since it gives us the optimum ST order to use for a given $\delta$. { This can be thought of as a change of coordinates and further arguments relating fig 1 to geometry can be found in \cite{supp2}.} An important point to clarify is the following. From the holographic results in \cite{Carmi:2017jqz}, it follows that the UV cutoff dependence should appear only along with the spatial volume dependence and not with the time dependence. In eq.(\ref{st56}), the UV cutoff dependence arises through the warp factor, but as it stands it is not clear if it affects both the spatial and time part (for small times, $\Omega\sim V t$). The gate counting argument clarifies what is happening. The reason $V$ needs to come with a UV cutoff dependence is that, as discussed in the introduction, $V\propto m$ where $m$ is the number of simple terms in the Hamiltonian, a discrete quantity. This motivates the introduction of a lattice cutoff which discretizes $V$. However, for $d$ the situation is different. We divided $d$  into $N$ steps using $N=d/\Delta$ but after optimization, this $\Delta$ dependence dropped out. Hence, we conclude that the UV cutoff dependence appears only with the spatial volume, consistent with \cite{Carmi:2017jqz}.

{\it \noindent Penalty factor flows---} Crudely speaking, $\delta$ measures the amount of coarse-graining. If we considered $d$ and the penalty factors $p$ to be independent of $\delta$, this would mean 
 that as $\delta\rightarrow 0$, $N_{\mathrm{gates}}^{\mathrm{opt},(\mathrm{local})}\rightarrow \infty$ as one would expect. It is a legitimate question, however, to ask if we could attempt  to make $N_{\mathrm{gates}}^{\mathrm{opt},(\mathrm{local})}$ independent of tolerance by  making the penalty factor a function of $\delta$ via $k_{\mathrm{opt}}$,  $p\rightarrow p \,(k_{\mathrm{opt}}).$ After all, an experimentalist would not have access to an ever increasing set of gates! The question then arises as to what in this circumstance would be a good measure of complexity.

The possibility mentioned above leads to flow equations for the penalty factors since the only way $d$ can depend on $k$ or the tolerance is through $p$. This is potentially a useful way of using the penalty factors and, to distinguish from the case where there is no correlation between the parameters in the cost Hamiltonian and the error tolerance, we will call this ``renormalized circuit complexity''.
We demand that $N_{\mathrm{gates}}^{\mathrm{opt},(\mathrm{local})}$ is independent of $k^{\mathrm{opt}}$, so differentiating with respect to $k^{\mathrm{opt}}$ and setting it to zero gives the differential equation \cite{qfta1}
\begin{equation}\label{st6}
\frac{d \ln d(k^{\mathrm{opt}})}{dk^{\mathrm{opt}}}=-4\ln 5\,.
\end{equation}
which gives $d=d_0 \exp(-4\ln 5 \,k^{\mathrm{opt}})$. Here $d_0$ satisfies $N_{\mathrm{gates}}^{\mathrm{opt},(\mathrm{local})} =  h \Omega_0$, where $\Omega_0 = V d_0$. Recall that we are taking $dm/dk^{\mathrm{opt}}=0$, i.e., in a sense we are talking about a fixed point since the number of simple parts $m$ that the averaged hamiltonian $\bar H$ splits into does not change. One can also find $d$ in terms of $\delta$, but for $p(k^{\mathrm{opt}})$,  a general solution to the differential equation \eqref{st6} is rather hard--it would need explicit knowledge of $d$ as a function of $p$. Let us focus on the situation when $p$ can be large. Here we will assume that $d(p)\sim p^\alpha$ and consider two logical possibilities: $\alpha>0, \alpha<0$. Several examples are discussed in \cite{supp1}. Let us write $d = \tilde{d}_0 \, p^{\alpha}$. The $\alpha=0$ case will be similar to $\alpha<0$ since we can write $d= \tilde{d}_0+ \tilde{d}_1 p^{\alpha}$ here. Defining an effective coupling $g$ via $g= 1/\ln p$ and $k^{\mathrm{opt}} = \ln (\Lambda/\Lambda_0)$ where $\Lambda_0$ is some reference scale, the differential equation for $g$ reads
\be
\beta_P(g) = \Lambda \, \frac{d g}{d \Lambda} = \frac{4 \ln 5}{\alpha} \, g^2, ~~~~~~~ \alpha \neq 0\,, \label{st7}
\ee
where $\beta_P(g)$ can be termed as the ``flow function"  for the effective coupling $g$.  The sign of the flow function is solely determined by the sign of $\alpha$. The solution of this equation is well known from standard quantum field theory results \cite{book}. It follows that for $\alpha > 0$, the coupling is increasing with $k^{\mathrm{opt}}$ implying the corresponding theory is becoming harder,  while for $\alpha <0$, it is the reverse. 

\begin{figure}[h]
 \includegraphics[width=0.42\textwidth]{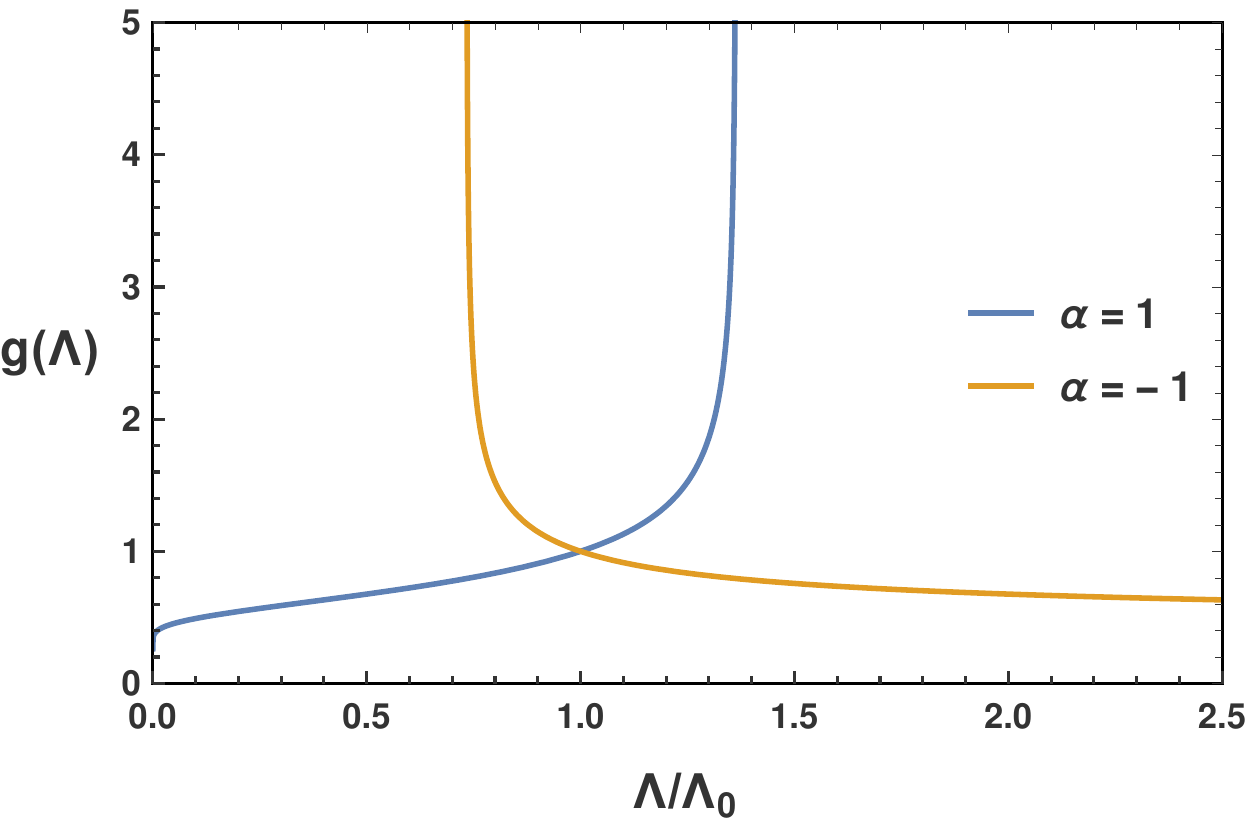}\\
\caption{
 Schematic flow of the coupling $g=1/\ln p$ with scale $k^{\mathrm{opt}}=\ln(\Lambda/\Lambda_0)$. Here $g\,(\Lambda=\Lambda_0) = 1$.} \label{alp1}
\end{figure}

The respective plots are shown in Fig.~(\ref{alp1}). Where the effective coupling $g$ blows up, the usage of penalty factors to suppress the hard gates, while keeping the total number of gates fixed, no longer helps--at this point one will need to switch to a dual description in terms of a different set of gates if available.  An important point to emphasize here is that we could have considered a penalty factor in front of any gate (which may be difficult to manufacture for instance); the flow equation is not restricted to penalty factors in front of interaction terms in the Hamiltonian.

Another point we emphasise is that whether we use the ST scheme as we have done here or some other scheme \cite{foot3} does not appear to be vital. We just need an exponential growth in the number of gates with $k$. If the growth was $e^{\gamma k}$ then the RHS of eq.(6) would be replaced by $-2\gamma$. Now notice that if we were to find the explicit circuit, there would still be some work to do. First, we need to solve the geodesic equation which gives $d$ as a function of $p$. Then we have $d$ as a function of $\delta$ by solving eq.(\ref{st6}) and using the transcendental equation \eqref{krel2}. 
 $d$ as a function of $\delta$ is monotonically increasing which would appear to be counterintuitive. But recall that this is because we are demanding that $N_{\mathrm{gates}}^{\mathrm {opt},(\mathrm{local})}$ is independent of $\delta$. At the same time, intuitively we would expect the circuit to become harder as the tolerance decreases. Then what is a good measure of the hardness of the circuit? First notice that $N_{\mathrm {gates}}^{\mathrm {opt},(\mathrm{local})}/\Omega\propto \exp \, (4\ln5\, k^{\mathrm {opt}})$ which monotonically increases with $k^{\mathrm {opt}}$ and hence with $1/\delta$. This in fact is true irrespective of whether we choose to hold the total number of gates fixed or not as can be easily seen from eq.(4). $N_{\mathrm{gates}}^{\mathrm{opt},(\mathrm{local})}/\Omega$ can be thought of as a density of gates.   We can also understand this by thinking of the total cost as given by the total circuit time cost and the total memory cost used in the computation \cite{presslec}. In this language, $d$ is the circuit time (also called circuit depth) while $ N_{\mathrm{gates}}^{\mathrm{opt},(\mathrm{local})}/d(\delta)$ quantifies the memory (space) needed (also called circuit width). We are keeping total cost fixed by decreasing the time cost while increasing the memory cost. We will argue that $A(\delta)\equiv \frac{1}{2}\ln \Big(N_{\mathrm{gates}}^{\mathrm{opt},(\mathrm{local})}/\Omega\Big)$ is connected to holographic $c$-theorems (eg. \cite{cthm}). 
 
 We can also correlate $g$ with binding complexity introduced \cite{JM13}. This counts only entangling gates. In our notation, this works out to be ${\mathcal C}_b=d/p$ where $p$ is taken to be very large and the penalty factor is associated with entangling gates. We could just use the same idea to count the hard gates. For $|\alpha|<1$, ${\mathcal C}_b$ and the effective coupling $g$ have the same trend with respect to $k$. Hence the effective coupling introduced before can be thought to be measuring binding complexity and for $0<\alpha<1$ increases as a function of $1/\delta$ which bears out the intuition that the circuit should become harder as tolerance decreases.

{\it \noindent Relation with AdS/CFT---} Let us now make some observations about how our description ties up with the AdS/CFT correspondence. In \cite{Cap1, Cap2,Cap3,Cap4,Cap5, Cap6} a definition of complexity (for $1+1$ dimensions) of has been proposed based on the path integral optimization technique. The complexity functional is the Liouville action. Furthermore, in  \cite{Camargo} inspired by the cMERA construction \cite{Vidal1,Vidal2},  it has been argued that the for a certain type of operator, one can obtain a Liouville type action as complexity functional at the leading order in a derivative expansion.  We start from the following action \cite{Cap6, Camargo},
 \begin{align}
    F_{\mathrm{holo}}\propto \int_{-\infty}^{-\epsilon} dt \int_{-\infty}^{\infty} dx \Big[ 2e^{2\phi} + p(\epsilon)^2 \Big((\partial_{t}\phi)^2 &+(\partial_x\phi)^2\Big) \nonumber \\
    &+\cdots\Big]. \label{act}
 \end{align}
 $p(\epsilon)^2$ is the penalty factor to give higher cost to gradients. Extremizing this action w.r.t $\phi$ (with the boundary condition $e^{2\phi(t=-\epsilon, x)}=2p(\epsilon)^2 /\epsilon^2$ \cite{Cap1,Cap2, Cap4, Cap5,Cap6} we get, 
 $e^{2\phi(t,x)}= 2\,p(\epsilon)^2 /|t|^2.$ This corresponds to the complexity of the ground-state \cite{Cap3}. Then evaluating $F_{\mathrm{holo}}$ on this solution and minimizing  further w.r.t $\epsilon$ we get (to make the total number of gates independent of $\epsilon$ similar to what we have done  previously),
\be \label{mon}
\frac{d}{d\epsilon} \Big(\frac{p(\epsilon)^2}{\epsilon}\Big)=0.
\ee
This can be solved using $p(\epsilon)\propto \sqrt{\epsilon}$.
 Defining the effective coupling, $g=1 /\ln p(\epsilon)$  and identifying $\epsilon=1/\Lambda$ as the UV cut-off  we get
 $
 \Lambda\frac{d g}{d\Lambda}=\frac{g^2}{2}.
 $
 Here we find a positive flow function for the penalty factor $p(\Lambda),$ suggesting the fact that the bulk circuit is easier while the boundary circuit is harder. \par
 Furthermore, we compute the on-shell Hamiltonian density ($h_{\mathrm{holo}}$) at $t=\epsilon$ corresponding to the action (\ref{act}):
 \be
 h_{\mathrm{holo}}=\frac{c}{8\pi}\frac{p(\epsilon)^2}{\epsilon^2}\,,
 \ee 
 where $c$ is the central charge.
 Now using the solution from \eqref{mon} we can easily see that,
 \begin{align}
& \frac{d h_{\mathrm{holo}}}{d\epsilon}< 0, \qquad
  \frac{d^2 h_{\mathrm{holo}}}{d\epsilon^2} >0. \label{ctheor}
 \end{align}
 Thus the Hamiltonian density evaluated at $t=\epsilon$ is a monotonically decreasing quantity in $\epsilon$. Note that \eqref{mon} which was the analog of $dN_{\mathrm{gates}}^{\mathrm{opt},(\mathrm{local})}/dk^{\mathrm{opt}}=0$ was vital in reaching this conclusion.  Now from (\ref{ctheor}), identifying $2  p(\epsilon)^2/\epsilon^2$ with $N_{\mathrm{gates}}^{\mathrm{opt},(\mathrm{local})}/Vd(\delta),$  we see that $A(\delta)\equiv \frac{1}{2}\ln \Big(N_{\mathrm{gates}}^{\mathrm{opt},(\mathrm{local})}/Vd(\delta)\Big)$ is also monotonic.  
 It is tempting to think that the monotonicity discussed above is connected to c-theorems in QFTs \cite{Zamolodchikov, cthm}. We will now establish a connection with holographic c-theorems following \cite{cthm}. In holographic c-theorems, the RG flow metric for a QFT living in $D$-dimensions is written as
\begin{equation}
ds^2=dr^2+e^{2A(r)}(-dt^2+d\vec{x}_{D-1}^2) \,.
\end{equation}
For Einstein gravity in the bulk, when the matter sector inducing the flow satisfies the null energy condition, 
$
a(r)=\frac{\pi^{\frac{D}{2}}}{\Gamma(\frac{D}{2})(A'(r))^{D-1}}\,,
$
can be shown to be monotonic $a'(r)\geq 0$.  At the fixed points $A(r)\propto r$ and we have an AdS metric. To connect with the previous discussion, we need $D=2$ and it will be convenient to make a change of coordinates $r=-\ln z$. In terms of this, it is easy to see that we must have $e^{2 A(z=\epsilon)}=2 \, p(\epsilon)^2 / \epsilon^2$. This also follows by realizing that the on-shell Liouville field is related to the warp factor \cite{Cap6}. Thus the density of gates is related to the geometric RG flow function $A(r)$ and the monotonicity in $h_{\rm{holo}}$ that we found is related to the monotonicity in $A(r)$. It is easy to check using \eqref{mon} that $A(r)\propto r$ in the $r$ coordinate. However, more generally in \eqref{mon} we should have matter contribution on the rhs. In such a circumstance, the fact that $A(r)$ should be monotonic in $r$ was argued in \cite{cthm} using the null energy condition.  In fact, it can be shown that to have $a'(r)>0$ one needs to put in matter satisfying the null energy condition to drive the flow. To model this using circuits, we would need to consider $m$ that changes with $k$. After optimization, this would lead to $N_{\mathrm{gates}}^{\mathrm{opt},(\mathrm{local})}\propto \exp[\gamma f(k^{\mathrm{opt}})]$ where $f$ is no longer linear in $k^{\mathrm{opt}}$.
 
{\it \noindent Discussion---} 
 In this paper, we have proposed a modification to Nielsen's circuit complexity calculation by introducing the Suzuki-Trotter iteration giving rise to what we call ``renormalized circuit complexity''.  First, we showed that the optimized gate counting leads to a linear dependence on the geodesic length and volume as suggested by holographic calculations unlike the cubic dependence found in \cite{Niel1}. While this is true for $g$-local Hamiltonians, the scaling will change for more general cases as was anticipated in \cite{sussrec}. The final form of the optimized gates $N_{\mathrm{gates}}^{\mathrm{opt},(\mathrm{local})}\sim \Omega\, h \exp(\gamma\, k^{\mathrm{opt}})$ appears to be universal for any iteration scheme; an unsolved question which we hope to return to in the near future is to prove that optimization cannot lead to sub-linear scaling with $d$ in any quantum algorithm. We found that $k^{\mathrm{opt}}$ is related to the tolerance hinting at an obvious connection with holography similar in spirit to the connection between holography and cMERA \cite{CT}. We further proposed using penalty factors to make the total number of gates independent of tolerance thereby leading to flow equations for the penalty factors. This picture also suggested that the density of gates  is a monotonically increasing function with $k^{\mathrm{opt}}$. The same physics arises from holography via the recent discussions on path-integral optimization \cite{Cap2,Cap6,Camargo} leading to the Hamiltonian density of the Liouville action playing the role of the monotonic flow function, which we further correlated with holographic c-theorems \cite{cthm}. Since there have been recent experimental realizations of the $3$-site spinless Hubbard model \cite{Thesis} and a proposal for realizing AdS/CFT  using quantum circuits \cite{Thesis1}, it will be very interesting to write down efficient circuits in these cases using the ideas in this paper. 

\begin{acknowledgements}
\noindent
{\sl Acknowledgements}: We thank Kausik Ghosh, Apoorva Patel, Tadashi Takayanagi  and Barry Sanders for useful discussions. We thank Barry Sanders and Tadashi Takayanagi  for comments on this draft. A.S. acknowledges support from a DST Swarnajayanti Fellowship Award DST/SJF/PSA-01/2013-14.  A.B. is supported by JSPS Grant-in-Aid for JSPS fellows (17F17023).
\end{acknowledgements}

  
  \onecolumngrid
\section*{
Supplementary Material}

\renewcommand{\thesection}{\Roman{section}}
\renewcommand{\thesubsection}{\thesection.\arabic{subsection}}
\renewcommand{\thesubsubsection}{\thesubsection.\arabic{subsubsection}}

\section{Derivation of $N_{\mathrm{gates}}$ and connection with geometry}
With the modification mentioned in eq. (1) of the main text, the total error becomes
\begin{align}
|| U - U_A || ~\leq~ \chi \frac{\mathcal{N} d}{p} + \frac{9}{2} \frac{d}{ \Delta}(m \Delta^2)
+ \frac{d}{\Delta} \frac{2 \, \kappa_m (2h \, 5^{k-1} \Delta)^{2k+1}}{r^{2k}}  \,.\label{st3}
\end{align}
where $\kappa_m$ is defined in the main text. Choosing $\Delta = \delta/m d$, one can make the second term $O(\delta)$. For the first and the third term, we slice the time interval by $r$ such that the first term and the third term together give the error of $O(\delta)$,
\begin{align}
\chi \frac{\mathcal{N} d}{p} + \frac{d}{\Delta} \frac{2 \, \kappa_m (2h \, 5^{k-1} \Delta)^{2k+1}}{r^{2k}} = \delta
\end{align}
so that for the full circuit, we have $|| U - U_A || \leq O(\delta)$. Thus we have,
\begin{align}
r = \Bigg\lceil 2 h \Delta \, 5^{k - \frac{1}{2}} \bigg(\frac{4 \, h \, d \, \kappa_m}{5 \,\tilde \delta}\bigg)^{\frac{1}{2k}} \Bigg\rceil, \label{st4}
\end{align}
where $\lceil ~ \rceil$ is the ceiling function and $\tilde{\delta} = \delta - \chi \mathcal{N} d/p$. Instead of working with the projected Hamiltonian we can of course choose to work with the full Hamiltonian itself and in that case we set $\chi =0$. In what follows we set $\chi = 0$. Now using the value of $r$ the total number of gates becomes $N_{\mathrm{gates}} = 2 m 5^{k-1} r d/\Delta$, which is given by
\begin{align}
N_{\mathrm{gates}} = O \bigg[h  m \, 5^{2k} d^{1+\frac{1}{2k}}\bigg(\frac{4\, h \,\kappa_m}{5 \, \delta}\bigg)^{\frac{1}{2k}} \bigg] \label{st5a}\,,
\end{align}
which is eq.(2) of the main text.

After the optimization is done following the reasoning in the main text, for a given error tolerance $\delta$, we will have a circuit similar to the one shown in fig 1 of the main text where any circuit beyond the $k^{\mathrm{opt}}$ corresponding to this error tolerance will satisfy $||U-U_A||\leq O(\delta)$. Fig 1 corresponds to a discretized geometry. $k^{\mathrm{opt}}$ corresponds to the radial direction. In eq. (5) in the main text, we could absorb the $\Omega$ dependence into $\delta$--the redefined $\delta$ can be thought of as the error tolerance per gate. Furthermore, for a fixed radius the total number of gates is given in eq.(4). We can distribute the total number of gates uniformly by demanding that 
$$ \int d^{n}x \sqrt{g}=O[\Omega \exp(4 \ln 5 k^{\mathrm{opt}})].$$ This will be the case if we write the fixed $k^{\mathrm{opt}}$ metric as $ds^2=\exp(8\ln 5 k^{\mathrm{opt}}/n) dx_i dx_i$, where $i=1,2,\cdots n$.  With this, the $\Omega$ factor will emerge naturally for a circuit time $d$. Here we have assumed that the Hamiltonian simulation is that of a Poincare invariant theory. Then in the continuum limit, introducing the dimensionful radial coordinate $r=k^{\mathrm{opt}}L$, we can think of the metric describing fig 1. as $ds^2=dr^2+\exp(8\ln 5 r/n L) (-dt^2+dx_i dx_i)$ in the scale and Poincare invariant situation--we have put the same warp factor in front of $-dt^2$ assuming Poincare invariance. This is the $AdS_{n+1}$ metric with the AdS radius set by $L n/(4\ln 5)$.

\section{Examples}
We now provide that some explicit examples for $d(p).$ As we have pointed out in the main text, penalty factors can be associated with any gate that is difficult to make.
\subsection {2-Majorana ($N=2$) SYK like model:} 
The Hamiltonian for this case,
\be
H=J_{1} \gamma_1+J_2 \gamma_2+J_3\gamma_1\gamma_2,
\ee
where, $\gamma_1, \gamma_2$ are two Majorana operators and $J_1, J_2, J_3$ are the random couplings. Following the analysis of \cite{JM14} if we suppress the contribution of the $\gamma_1\gamma_2$ by a large penalty factor, then we can show that in this case $d(p) \sim p^0.$ This means that as $k$ becomes large, the circuit will involve less of these gates.

\subsection{Complexity for  ground state of free scalar field theory:} 

We compute the complexity for the ground state of a free scalar field theory in $1+1$ dimensions. We  discretize it on the lattice.  Effectively we get a system of coupled oscillators. The Hamiltonian is \cite{JM},
\be
H=\frac{1}{2}\sum_{i=0}^{N-1}\Big(p_i^2+\omega^2 x_i^2+\Omega^2(x_i-x_{i+1})^2\Big),
\ee
where, $\omega=m, \Omega=\frac{1}{\delta}.$ $\delta$ is the lattice spacing and $i$ denotes the position on the lattice. For simplicty we focus on the 2 coupled oscillator. For this case the Hamiltonian becomes,
\be
H=\frac{1}{2}(p_1^2+p_2^2+(\omega^2+\Omega^2)(x_1^2+x_2^2)+2\Omega^2 x_1 x_2)
\ee

The ground state for this Hamiltonian is given by,
\be
\psi^T(x_1,x_2)=\frac{(\omega_1\omega_2-\beta^2)^{1/4}}{\sqrt{\pi}}\exp\Big(-\frac{\omega_1}{2} x_1^2 -\frac{\omega_2}{2}x_2^2-\beta  x_1 x_2\Big),
\ee
 with, 
 \be
 \omega_1=\omega_2=\frac{1}{2}(\omega+\sqrt{\omega^2+2\Omega^2}),\beta=\frac{1}{2}(\omega-\sqrt{\omega^2+2\Omega^2}).
 \ee
  
We will compute the complexity of this state w.r.t to the following state with no entanglement between $x_1$ and $x_2$.
 \be
 \psi^R(x_1,x_2)=\sqrt{\frac{\omega_r}{\pi}}\exp\Big(-\frac{\omega_r}{2}(x_1^2+x_2^2)\Big),
 \ee
 where $\omega_r$ is a reference frequency. Now we have to construct the optimal circuit which will take us from $\psi^R(x_1,x_2)$ to $\psi^T(x_1,x_2). $ 
 \be \label{eq1}
 \psi^T(x_1,x_2)= \overleftarrow{\mathcal{P}}\exp\Big(\int_{0}^{1} Y^{I}(s)M_{I} ds\Big)\psi^R(x_1,x_2),
 \ee
 where the circuit $U(s)= \overleftarrow{\mathcal{P}}\exp\Big(\int_{0}^{s} Y^{I}(s)M_{I} ds\Big)$ is generated by the four generators  ($M_I$)  of $GL(2, R)$ and $Y^{I}$ are some control functions. $s$ paratmetrize the path in the space of circuit and for $s=0$ we have $U(s=0)=I,$ we get back the reference state.  For $s=1$ we get the target state. We need to optimize $Y^{I}$  as a function of $s$ to get the optimal circuit. This is achieved by first writing down an action for $Y^{I}$ and then minimizing it. We choose the following functional,
 \be \label{cost}
 C(U(s))=\int_{ 0}^1 ds \sqrt{\sum_{I,J}g_{I J} Y^{I} (s)Y^{J}(s)}. 
 \ee
$g_{IJ}$ are the penalty factors. For our case, the generators of the circuit takes the following form, $M_{ab}=(i x_a p_b+\frac{1}{2}\delta_{ab}),$ where the index $I$ which appears in eq.(\ref{eq1}),  $\in \{11,12,21,22 \}.$  Now $M_{11}$ and $M_{22}$ both corresponds to scaling generators which scales the coefficients of $x_1$ and $x_2$ and $M_{12}$ and $M_{21}$ are the entangling generators which shifts $x_1$ by $x_2$ and vice versa thereby generating $x_1 x_2$ term in the wavefunction. Both $\psi^{T}(x_1,x_2)$ and $\psi^{R}(x_1,x_2)$ can be written as \be \psi(x_1,x_2)=\mathcal{N}\exp\Big(x_a. A_{ab}. x_b\Big),\ee
where, $\vec x=\{x_1,x_2\}.$ Given this basis vector, the generators $M'$s take the form of a $2 \times 2$ matrix.   
\begin{align}\begin{split}
M_{11}=\left(
\begin{array}{cc}
 1 & 0 \\
 0 & 0 \\
\end{array}
\right), M_{22}=\left(
\begin{array}{cc}
 0 & 0 \\
 0 & 1 \\
\end{array}
\right),M_{12}=\left(
\begin{array}{cc}
 0 & 1 \\
 0 & 0 \\
\end{array}
\right), M_{21}=\left(
\begin{array}{cc}
 0 & 0 \\
 1 & 0 \\
\end{array}
\right). \label{gen}
\end{split}
\end{align}
These are nothing but the generators of $GL(2, R).$ Following \cite{JM} we can conveniently parametrize $U(s)$ in the following way,
 \begin{align}
\begin{split} \label{unit}
U(s)= \exp(y_3(s))\left(
\begin{array}{cc}
 x_{0}- x_{3} &  x_{2}- x_{1} \\
  x_{2}+ x_{1} &  x_{0}+ x_{3} , \\
\end{array}
\right),
\end{split}
\end{align}
with,
\begin{align}
\begin{split}
&  x_0=\cos(\tau(s))\cosh(\rho(s)), x_1=\sin(\tau(s))\cosh(\rho(s)),\\&
x_2=\cos(\theta(s))\sinh(\rho(s)), x_3=\sin(\theta(s))\sinh(\rho(s)).
\end{split}
\end{align}
\subsection*{Penalize entangling gates}
First we will penalize the entangling gates corresponding to the generators $M_{12}$ and $M_{21}.$ We set following \cite{JM},
\be \label{pen}
g_{IJ}=\rm{diag}\{1, p^2,p^2, 1\}.
\ee
Given this and using (\ref{unit}), the complexity functional (\ref{cost}) becomes a distance functional 
\be d(p)=\sqrt{2}\int_0^1 ds\, k,
\ee
where,
\begin{align}
\begin{split}
k^2= &\dot y^2+(p^2-(p^2-1) \sin^2(2 x))\dot \rho^2- (p^2-1)\sin(4 x)\sinh(2\rho)\dot \rho\dot z\\&+ p^2 \dot x^2+ (p^2\cosh(4\rho)-(p^2-1)\cos^2(2x)\sinh^2(2\rho))\dot z^2-2 p^2\cosh(2\rho)\dot x \dot z.
\end{split}
\end{align}
for the manifold associated with the following metric,
\begin{align}
ds^2= &2 dy^2+ 2 (p^2-(p^2-1)\sin^2(2 x))d\rho^2- 2(p^2-1) \sin(4 x)\sinh(2\rho) d\rho dz\\&+ 2p^2 dx^2+2 (p^2 \cosh(4\rho)-(p^2-1)\cos^2(2 x) \sinh^2(2\rho))dz^2- 4p^2 \cosh(2\rho) dx dz.
\end{align}
Here the dot denote the derivative w.r.t $s$ and $\theta=x+z, \tau=x-z.$\\
Then following the analysis of \cite{JM} one can show that for large $p,$ 
\be \label{result1}
d(p) \approx p+\cdots  
\ee
\subsection*{Penalize scaling gates}
First we will penalize the entangling gates corresponding to the generators $M_{11}$ and $M_{22}.$ We set,
\be \label{pen1}
g_{IJ}=\rm{diag}\{p^2,1,1, p^2\}.
\ee
In this case, the complexity functional (\ref{cost}) becomes a distance functional 
\be d(p)=\sqrt{\frac{1}{2}}\int_0^1 ds\, k,
\ee
where,
\begin{align}
\begin{split}
k^2= &\Big(2  \left(p^2+1-\left(p^2-1\right) \cos (4 x)\right) \dot \rho^2+4 \left(p^2-1\right)\sinh (2 \rho) \sin (4 x) \dot \rho \dot z \\&+ \left(\left(p^2+3\right) \cosh (4 \rho )+\left(p^2-1\right) \left(2 \sinh ^2(2 \rho) \cos (4 x)-1\right)\right) \dot z^2+4 p^2 \dot y^2\\&-8\, \cosh (2 \rho) \dot x \dot z+4\, \dot x^2\Big),
\end{split}
\end{align}
for the manifold associated with the following metric,
\begin{align}
ds^2=& \frac{1}{2} \Big(2  \left(p^2+1-\left(p^2-1\right) \cos (4 x)\right)  d\rho^2+4 \left(p^2-1\right)\sinh (2 \rho) \sin (4 x) d\rho\, dz \\&+ \left(\left(p^2+3\right) \cosh (4 \rho )+\left(p^2-1\right) \left(2 \sinh ^2(2 \rho) \cos (4 x)-1\right)\right)  dz^2+4 p^2  dy^2\\&-8\, \cosh (2 \rho)  dx dz+4\, dx^2\Big).
\end{align}

Performing an analysis  similar to one done in \cite{JM} we can show that in the large $p$ limit,
\be \label{result2}
d(p)\approx p^0+\cdots
\ee  

The results in (\ref{result1}) and (\ref{result2}) can be intuitively explained in the following way. If we look the algebra of generators  we can see that, 
\be
[M_{12}, M_{21}]=M_{11}-M_{22}.
\ee
The commutator of the entangling gates generate the scaling gates.  But the converse is not true as the scaling gates commutate with each other.  Here we have used (\ref{gen}).  So naturally if we suppress the  entangling gates we will be requiring more number of gates to reproduce the target state compared to the case when we suppress the scaling gates as the effects of the scaling can be still be generated by the entangling gates. This shows the plausibility of the results mentioned in (\ref{result1}) and (\ref{result2}).

\subsection{Interacting scalar field theory:}  We consider first $\lambda \phi^4$ theory in $D+1$ dimensions. A first principle analysis will require working out the algebra of operators systematically and computing the geodesic--this has not been done and appears difficult with current technology \cite{Cotler2, Cotler3} \footnote{One can alternatively, consider an approximated wavefunction which is of Gaussian in nature as argued in \cite{Gapp,Gapp1,Gapp2,Gapp3,Gapp4,Gapp5}.}. However, we can give a heuristic argument as follows. Consider the discretized theory on the lattice. Then following the analysis of \cite{AB}, we have the perturbative term $(\delta\, p)^{4-D}$ suppressing the contribution of the non-gaussian operators in the expression for complexity\footnote{In \cite{AB}, the analysis was for state complexity which can be thought of as finding the operator which takes the initial state to final state and which leads to the lowest complexity.}. Again, $\delta$ is the lattice spacing and $\delta \rightarrow 0$ to recover the continuous theory. We demand that the result is perturbatively finite once (large) penalty factors are included. So for $ D < 4 $ we have $\alpha > 0$ and for $D > 4$ we have $\alpha < 0. $ This is consistent with the fact that there is a Wilson-Fisher fixed point for $D<4$ in the IR since an efficient description in this case will need a lower value of $k$. For $D>4$, one can efficiently describe using a large $k$ and a large $p$ which would correspond to the Gaussian fixed point. A similar argument can be given for the $\phi^3$ theory as well. It will be gratifying to have a more rigorous argument based on the determination of $\alpha$ in a non-perturbative framework.

\subsection{Complexity for  time evolution operator: a perturbative computation}

We compute the circuit complexity  for the time evolution operator $U=e^{-i H t },$ where $H$ is the Hamiltonian. We want to find the efficient circuit which represents this the unitary.  We essentially follow the steps as before.  The circuit is parametrized as,
\be
U(s)= \overleftarrow{\mathcal{P}}\exp\Big(-i\,\int_{0}^{s} Y^{I}(s)M_{I} ds\Big) 
\ee
as before. Here $U(s=0)=I$ and $U(s=1)= e^{-i H t}.$ $I$ is the identity operator. Given this boundary condition we again proceed to to minimize the const functional (\ref{cost}).  Again after suitable parametrization (\ref{cost}) becomes the distance functional $d(p)$  for a certain manifold. Now we will consider $t$ is small. So it will enable us to do the calculation perturbatively in $t.$
 $H$ typically takes the following form,
\be \label{cond}
 H=\sum_{I} h^{I} M_{I}.
\ee
$M_{I}$ forms a complete basis. These operators typically satisfy the following Lie-algebra,
\be
[M_{I}, M_{J}]=i\sum_{K}f_{IJ}{}^{K} M_{K},
\ee 
where, $f_{IJ}{}^{K}$ are the structure constants. Now we have to solve these $Y^{I}s$ by minimizing (\ref{cost}). As $t$ is small, $Y^{I}s$  can be solved perturbatively in $t.$ We quote simply the  results  here. The detailed calculations are done in \cite{JM13}. 
\be \label{expand}
Y^{I}(s)=t\, h^{I}-\frac{1}{2} t^2(1-2 s)\sum_{J,K} C_{J K}{}^{I} h^{J} h^{K}+\mathcal{O}(t^3)+\cdots,
\ee
where, $h^{I}s$ are defined in (\ref{cond}) and 
\be
C_{J K}{}^{I}=\sum f_{J L}{}^{M} (\mathcal{I}^{-1})^I_{M} \mathcal{I}^{L}_{K}, \quad \mathcal{I}^{I}_{J}=\sum_{M} K^{I M} g_{M J},\quad  K_{IJ}=\sum_{M, L} f_{I M}{}^{L} f_{J L}{}^{M}.
\ee
Also we have used the fact that,
\be
g_{IJ}=\frac{c_{I}+c_{J}}{2} K_{IJ}
\ee
Again we can penalize certain gates by choose $c_{I} = p^2$ for some $I$ and $p$ is very large. For other gates we can set $c_{I}=1.$ We can easily see that in our previous example,  $K_{IJ}=\delta_{IJ}$ and we can get either (\ref{pen}) or (\ref{pen1}) depending on whether we choose to suppress the entangling or scaling gates. Now using (\ref{expand}) we get after evaluating (\ref{cost}),
\be \label{finansw}
d(p)=t\, \sqrt{\sum_{I,J}g_{I J} h^{I}h^{J}}+\mathcal{O}(t^2)+\cdots
\ee
 
Now from (\ref{finansw})  it is evident that, for large $p$  depending on the structure of $g_{IJ}$ we will have the following leading order behaviour of $d(p).$
\be
d(p)\approx  t\, p^{\alpha}+\mathcal{O}(t^2)+\cdots,
\ee
where either $\alpha=0$ or $1.$



\begin{thebibliography}{12}

\bibitem{Niel1}
M.~A. Nielsen, M.~R. Dowling, M.~Gu, and A.~C. Doherty, 
 {\em Science} {\bf 311} (2006) 1133--1135,
  [\href{http://arxiv.org/abs/quant-ph/0603161}{{\tt quant-ph/0603161}}].


\bibitem{Niel2}
M.~ A.~ Nielsen, M.~ R.~Dowling, M.~ Gu, and A.~ C. Doherty,
Phys.~Rev.~A.~73,~062323 [arXiv:quant-ph/0603160].

\bibitem{Niel3}
M.~A. Nielsen,  
 Quantum Information and Computation, 6, 213 (2006), arXiv:0502070[quant-ph].

\bibitem{Niel4}
M.~A. Nielsen and M.~R. Dowling, 
Quantum Information \& Computation, 8, 861 (2008) quant-ph/0701004.





\bibitem{JM} 
  R.~Jefferson and R.~C.~Myers,
  JHEP {\bf 1710}, 107 (2017),
  arXiv:1707.08570 [hep-th].
 
 \bibitem{JM2a} 
  S.~Chapman, M.~P.~Heller, H.~Marrochio and F.~Pastawski,
  Phys.\ Rev.\ Lett.\  {\bf 120}, no. 12, 121602 (2018),
  arXiv:1707.08582 [hep-th].
 

 \bibitem{JM10} 
  S.~Chapman, J.~Eisert, L.~Hackl, M.~P.~Heller, R.~Jefferson, H.~Marrochio and R.~C.~Myers,
  SciPost Phys.\  {\bf 6}, no. 3, 034 (2019),
 arXiv:1810.05151 [hep-th].

\bibitem{JM1} 
  L.~Hackl and R.~C.~Myers,
 `
  JHEP {\bf 1807}, 139 (2018)
   [arXiv:1803.10638 [hep-th]].
 


\bibitem{JM2} 
  M.~Guo, J.~Hernandez, R.~C.~Myers and S.~M.~Ruan,
  JHEP {\bf 1810}, 011 (2018)
 [arXiv:1807.07677 [hep-th]].


 

 \bibitem{JM3}
R.~Q.~Yang,
  Phys.\ Rev.\ D {\bf 97} (2018) no.6,  066004
    [arXiv:1709.00921 [hep-th]].

  


 

\bibitem{JM4} 
  R.~Khan, C.~Krishnan and S.~Sharma,
  Phys.\ Rev.\ D {\bf 98}, no. 12, 126001 (2018)
  [arXiv:1801.07620 [hep-th]].
  

\bibitem{JM5} 
  R.~Q.~Yang, Y.~S.~An, C.~Niu, C.~Y.~Zhang and K.~Y.~Kim,
  Eur.\ Phys.\ J.\ C {\bf 79}, no. 2, 109 (2019)
  [arXiv:1803.01797 [hep-th]].
  


 

\bibitem{JM6}
  D.~W.~F.~Alves and G.~Camilo,
  JHEP {\bf 1806} (2018) 029
  [arXiv:1804.00107 [hep-th]].
  
  \bibitem{JM7}
  P.~Caputa and J.~M.~Magan,
  Phys.\ Rev.\ Lett.\  {\bf 122}  no.23,  231302 (2019)
  [arXiv:1807.04422 [hep-th]].


\bibitem{JM8} 
  H.~A.~Camargo, P.~Caputa, D.~Das, M.~P.~Heller and R.~Jefferson,
  Phys.\ Rev.\ Lett.\  {\bf 122}, no. 8, 081601 (2019)
  [arXiv:1807.07075 [hep-th]].
 
\bibitem{JM8a}
  W.~Chemissany and T.~J.~Osborne,
  JHEP {\bf 1612} (2016) 055
[arXiv:1605.07768 [hep-th]].


 \bibitem{JM9a} 
  A.~R.~Brown and L.~Susskind,
  Phys.\ Rev.\ D {\bf 97}, no. 8, 086015 (2018)
   [arXiv:1701.01107 [hep-th]].
 
 \bibitem{JM9}
 T.~Ali, A.~Bhattacharyya, S.~Shajidul Haque, E.~H.~Kim and N.~Moynihan,
  JHEP {\bf 1904} (2019) 087
  [arXiv:1810.02734 [hep-th]].
  


 
  \bibitem{JM11} 
 R.~Q.~Yang and K.~Y.~Kim,
  JHEP {\bf 1903}, 010 (2019)
   [arXiv:1810.09405 [hep-th]].
   
   \bibitem{JM12}
T.~Ali, A.~Bhattacharyya, S.~Shajidul Haque, E.~H.~Kim and N.~Moynihan,
  arXiv:1811.05985 [hep-th].



  \bibitem{JM13a}
  A.~Bernamonti, F.~Galli, J.~Hernandez, R.~C.~Myers, S.~M.~Ruan and J.~Sim�n,
  arXiv:1903.04511 [hep-th].
 
 
  \bibitem{JM14}
  V.~Balasubramanian, M.~Decross, A.~Kar and O.~Parrikar,
  arXiv:1905.05765 [hep-th].

\bibitem{JM14a}
  I.~Akal,
  arXiv:1903.06156 [hep-th].


\bibitem{JM15} 
  T.~Ali, A.~Bhattacharyya, S.~S.~Haque, E.~H.~Kim, N.~Moynihan and J.~Murugan,
  arXiv:1905.13534 [hep-th].
   
\bibitem{JM16} 
  R.~Q.~Yang and K.~Y.~Kim,
  arXiv:1906.02052 [hep-th].\\
  R.~Q.~Yang, Y.~S.~An, C.~Niu, C.~Y.~Zhang and K.~Y.~Kim,
  arXiv:1906.02063 [hep-th].
  


 
\bibitem{sussrec} 
  A.~R.~Brown and L.~Susskind,
  Phys.\ Rev.\ D {\bf 100}, no. 4, 046020 (2019),
  arXiv:1903.12621 [hep-th].

  
  

\bibitem{Susskind4}
D.~Stanford and L.~Susskind, 
   Phys. Rev. {\bf D 90} (2014), no.~12 126007,
 arXiv:1406.2678 [hep-th].

\bibitem{Susskind5}
  A.~R.~Brown, D.~A.~Roberts, L.~Susskind, B.~Swingle and Y.~Zhao,
  Phys.\ Rev.\ Lett.\  {\bf 116} (2016) no.19,  191301, 
  arXiv:1509.07876 [hep-th].

 \bibitem{Susskind1}
 L.~Susskind,
  arXiv:1808.09941 [hep-th].



\bibitem{Hol5}
S.~Chapman, H.~Marrochio, and R.~C. Myers,
JHEP {\bf 01} (2017) 062,
arXiv:1610.08063[hep-th].

\bibitem{Carmi:2017jqz}
D. Carmi, S. Chapman, H. Marrochio, R. C. Myers, and S. Sugishita, 
JHEP 11, 188 (2017), arXiv:1709.10184[hep-th].


\bibitem{Hol6}
D.~Carmi, R.~C. Myers, and P.~Rath, 
    JHEP {\bf 03} (2017) 118,
  arXiv:1612.00433[hep-th].

  



  

  \bibitem{Susskind1a}
L.~Susskind, 
  Fortsch. Phys. {\bf 64} (2016) 24--43,
 arXiv:1403.5695[hep-th].




\bibitem{Susskind2}
L.~Susskind and Y.~Zhao, 
 arXiv:1408.2823 [hep-th].

\bibitem{Susskind3}
L.~Susskind, 
Fortsch. Phys. {\bf 64}
  (2016) 49--71,arXiv:1411.0690 [hep-th].




\bibitem{Susskind6}
A.~R. Brown, D.~A. Roberts, L.~Susskind, B.~Swingle, and Y.~Zhao, 
Phys. Rev. {\bf D93} (2016),
  no.~8 086006, arXiv:1512.04993 [hep-th].


\bibitem{Susskindlec} 
  L.~Susskind,
  arXiv:1810.11563 [hep-th].


\bibitem{Alishahiha:2015rta}
M.~Alishahiha, 
Phys. Rev. {\bf D92}, no.~12 126009  (2015), arXiv:1509.06614[hep-th].

\bibitem{Hol1}
R.-G. Cai, S.-M. Ruan, S.-J. Wang, R.-Q. Yang, and R.-H. Peng,
JHEP {\bf 09} (2016) 161, arXiv:1606.08307[hep-th].

\bibitem{Hol2}
A.~R. Brown, L.~Susskind, and Y.~Zhao, 
Phys. Rev. {\bf D95} (2017), no.~4 045010, arXiv:1608.02612 [hep-th].


\bibitem{Hol3}
L.~Lehner, R.~C. Myers, E.~Poisson, and R.~D. Sorkin, 
Phys. Rev. {\bf D94}, no.~8  084046  (2016),  arXiv:1609.00207[hep-th].


\bibitem{Hol7}
A.~Reynolds and S.~F. Ross, 
   Class. Quant. Grav. {\bf 34} (2017), no.~10 105004,
  arXiv:1612.05439 [hep-th].

\bibitem{Hol12}
D.~Carmi, S.~Chapman, H.~Marrochio, R.~C. Myers, and S.~Sugishita, 
JHEP {\bf 11} (2017) 188,
   arXiv:1709.10184 [hep-th].
   
\bibitem{Kim} 
  R.~Q.~Yang, C.~Niu, C.~Y.~Zhang and K.~Y.~Kim,
  JHEP {\bf 1802}, 082 (2018)
   [arXiv:1710.00600 [hep-th]].
 



\bibitem{Hol18} 
  M.~Moosa,
  JHEP {\bf 1803}, 031 (2018)
  [arXiv:1711.02668 [hep-th]].

\bibitem{Hol18a}
  M.~Moosa,
  Phys.\ Rev.\ D {\bf 97}, no.10,  106016 (2018)
  [arXiv:1712.07137 [hep-th]].


\bibitem{Hol21} 
  S.~Chapman, H.~Marrochio and R.~C.~Myers,
  JHEP {\bf 1806}, 046 (2018)
  [arXiv:1804.07410 [hep-th]].

\bibitem{Hol24} 
  S.~Chapman, H.~Marrochio and R.~C.~Myers,
  JHEP {\bf 1806}, 114 (2018)
  [arXiv:1805.07262 [hep-th]].

\bibitem{Flory}
  M.~Flory and N.~Miekley,
  JHEP {\bf 1905} (2019) 003
  [arXiv:1806.08376 [hep-th]].
  
  \bibitem{Flory1}
 M.~Flory,
  JHEP {\bf 1905} (2019) 086
  [arXiv:1902.06499 [hep-th]].

\bibitem{Flory2}
  M.~Flory,
  JHEP {\bf 1706} (2017) 131
  [arXiv:1702.06386 [hep-th]].

\bibitem{Hol26} 
 J.~Couch, S.~Eccles, T.~Jacobson and P.~Nguyen,
  JHEP {\bf 1811} (2018) 044
  [arXiv:1807.02186 [hep-th]].


   \bibitem{Hol27} 
  A.~R.~Brown, H.~Gharibyan, H.~W.~Lin, L.~Susskind, L.~Thorlacius and Y.~Zhao,
  Phys.\ Rev.\ D {\bf 99}, no. 4, 046016 (2019)
    [arXiv:1810.08741 [hep-th]].


\bibitem{Hol28} 
  K.~Goto, H.~Marrochio, R.~C.~Myers, L.~Queimada and B.~Yoshida,
  JHEP {\bf 1902}, 160 (2019)
  [arXiv:1901.00014 [hep-th]].
   
    
\bibitem{rub}
  J.~Jiang and B.~X.~Ge,
  Phys.\ Rev.\ D {\bf 99} (2019) no.12,  126006
  [arXiv:1905.08447 [hep-th]]



\bibitem{nielsenchuang}
 I.~ Chuang, and M.~ Nielsen, ``Quantum Computation and Quantum Information", Cambridge University Press, 2nd ed. (2010).


\bibitem{presslec2}
\url {http://www.theory.caltech.edu/~preskill/ph219/quantum-simulation-23feb2009.pdf}





\bibitem{Suzuki1}
M.~ Suzuki,
Physics Letters A, Volume 146, Issue 6, p. 319-323.

\bibitem{Suzuki2}
M.~ Suzuki,
Journal of Mathematical Physics, Volume 32, Issue 2, February 1991, pp.400-407.

\bibitem{Barry1a}
D.~ W. ~Berry, G~ Ahokas, R.~ Cleve, and B.~ C.~ Sanders,
chapter 4 of `` Mathematics of quantum computation and quantum technology", Chapman and Hall, pages 89-112 (2007)


\bibitem{Barry1}
D.~ W. ~Berry, G~ Ahokas, R.~ Cleve, and B.~ C.~ Sanders,
Communications in Mathematical Physics 270, 359 (2007).

\bibitem{Barry2}
N.~ Wiebe, D.~ W. Berry, P.~ Hoyer, and B.~ C. Sanders,
J. Phys. A: Math. Theor. 43, 065203 (2010).

\bibitem{foot1}
 $|| \cdot ||$ is the operator norm which is defined as $||X|| = \mathrm{max}_{\ket{\psi}} |\braket{\psi|X|\psi}|$, and the maximization is over all normalized vectors, $|\braket{\psi|\psi}|^2 = 1$, for any operator $X$.

\bibitem{supp2}
see section (I) of supplementary material for a detailed derivation which includes references \cite{JM14, JM, Cotler2, Cotler3, Arpan}.

\bibitem{com1}{The authors of    \cite{JM,JM2a,JM10,JM13} computed this geodesic distance for various systems. We denote it by $d(p)^{\mathrm{ref}}.$ In our notation the geodesic distance $d(p)$ is simply related to this $d(p)^{\mathrm{ref}}$ as $d(p)^{\mathrm{ref}}=d(p)V,$ where $V$ is the spatial volume, in the large volume limit. To see this, note that in these papers mentioned above, there is no splitting into $m$ parts in the calculation so the volume dependence comes entirely from $d$ there.}









\bibitem{supp}
see section (II.D) of supplementary material. 

\bibitem{supp1}
see section (II) of supplementary material.

\bibitem{book}
M.~ Srednicki, ``Quantum Field Theory", Cambridge University Press, Jan 25, 2007




\bibitem{foot3}
For example, the leap-frog algorithm in M.~ Creutz and A.~ Gocksch, 
Phys.\ Rev.\ Lett. 63, 9(1989), which has a similar recursion relation as the ST discussed here.

\bibitem{presslec}
\url{http://www.theory.caltech.edu/~preskill/ph219/chap5_13.pdf}


\bibitem{qfta1} The analogy with quantum field theory is to recall that in continuum RG, the beta function equations are derived by demanding that the bare action is independent of the fictitious RG scale.
  
  \bibitem{Zamolodchikov}
  A.~B.~Zamolodchikov,
  JETP Lett.\  {\bf 43} (1986) 730
   [Pisma Zh.\ Eksp.\ Teor.\ Fiz.\  {\bf 43} (1986) 565].

\bibitem{cthm}
 R.~C.~Myers and A.~Sinha,
  Phys.\ Rev.\ D {\bf 82}, 046006 (2010),
  arXiv:1006.1263 [hep-th].\\
 R.~C.~Myers and A.~Sinha,
  JHEP {\bf 1101}, 125 (2011),
    arXiv:1011.5819 [hep-th].

\bibitem{JM13}
  V.~Balasubramanian, M.~DeCross, A.~Kar and O.~Parrikar,
  JHEP {\bf 1902} (2019) 069,
   arXiv:1811.04085 [hep-th].

\bibitem{Cap1}
  P.~Caputa, N.~Kundu, M.~Miyaji, T.~Takayanagi and K.~Watanabe,
  Phys.\ Rev.\ Lett.\  {\bf 119} (2017) no.7,  071602,
  arXiv:1703.00456 [hep-th].

\bibitem{Cap2}
  P.~Caputa, N.~Kundu, M.~Miyaji, T.~Takayanagi and K.~Watanabe,
  JHEP {\bf 1711} (2017) 097,
  arXiv:1706.07056 [hep-th].

\bibitem{Cap3}
  B.~Czech,
  Phys.\ Rev.\ Lett.\  {\bf 120} (2018) no.3,  031601,
  arXiv:1706.00965 [hep-th].

\bibitem{Cap4}
 J.~Molina-Vilaplana and A.~Del Campo,
  JHEP {\bf 1808} (2018) 012
    [arXiv:1803.02356 [hep-th]].

\bibitem{Cap5}
  A.~Bhattacharyya, P.~Caputa, S.~R.~Das, N.~Kundu, M.~Miyaji and T.~Takayanagi,
  JHEP {\bf 1807} (2018) 086
    [arXiv:1804.01999 [hep-th]].


\bibitem{Cap6}
  T.~Takayanagi,
  JHEP {\bf 1812} (2018) 048,
    arXiv:1808.09072 [hep-th].


\bibitem{Camargo} 
  H.~A.~Camargo, M.~P.~Heller, R.~Jefferson and J.~Knaute,
  Phys.\ Rev.\ Lett.\  {\bf 123} (2019) no.1,  011601,
  arXiv:1904.02713 [hep-th].

  

\bibitem{Vidal1}
  A.~Milsted and G.~Vidal,
  arXiv:1807.02501 [cond-mat.str-el].


\bibitem{Vidal2}
  A.~Milsted and G.~Vidal,
  arXiv:1812.00529 [hep-th].

 \bibitem{CT} 
 M.~Nozaki, S.~Ryu and T.~Takayanagi,
  JHEP {\bf 1210} (2012) 193
  [arXiv:1208.3469 [hep-th]].




\bibitem{Thesis}
 R.~Barends, L.~Lamata, J.~Kelly, L.~Garc{\'\i}a-{\'A}lvarez, A.~G.~Fowler, A.~Megrant, E.~Jeffrey, T.~C.~White, D.~Sank and J.~Y.~Mutus, 
 Nature Communications 6, 7654 (2015),  arXiv:1501.07703 [quant-ph].


\bibitem{Thesis1}
 L.~Garc{\'\i}a-{\'A}lvarez, I.~L.~Egusquiza, L.~Lamata, A.~del Campo, J.~Sonner and E.~Solano,
  Phys.\ Rev.\ Lett.\  {\bf 119} (2017) no.4,  040501,
  arXiv:1607.08560 [quant-ph].



  
\bibitem{Cotler2}
J.~Cotler, M.~R.~Mohammadi Mozaffar, A.~Mollabashi and A.~Naseh,
Fortsch.\ Phys.\  {\bf 67}, no. 10, 1900038 (2019),
  arXiv:1806.02831 [hep-th].

\bibitem{Cotler3}
  J.~Cotler, M.~R.~Mohammadi Mozaffar, A.~Mollabashi and A.~Naseh,
  Phys.\ Rev.\ D {\bf 99}, no. 8, 085005 (2019)
  arXiv:1806.02835 [hep-th].

\bibitem{Arpan} 
  A.~Bhattacharyya, A.~Shekar and A.~Sinha,
  JHEP {\bf 1810}, 140 (2018),
  arXiv:1808.03105 [hep-th].
 
  

  










\end{thebibliography}

\begin{thebibliography}{}


\bibitem{JM14}
  V.~Balasubramanian, M.~Decross, A.~Kar and O.~Parrikar,
  JHEP {\bf 2001}, 134 (2020),
  arXiv:1905.05765 [hep-th].

\bibitem{JM}
R.~A. Jefferson and R.~C. Myers, 
JHEP {\bf 10} (2017) 107, 
arXiv:1707.08570[hep-th].
  

\bibitem{Cotler2}
J.~Cotler, M.~R.~Mohammadi Mozaffar, A.~Mollabashi and A.~Naseh,
Fortsch.\ Phys.\  {\bf 67}, no. 10, 1900038 (2019),
  arXiv:1806.02831 [hep-th].


\bibitem{Cotler3}
  J.~Cotler, M.~R.~Mohammadi Mozaffar, A.~Mollabashi and A.~Naseh,
  Phys.\ Rev.\ D {\bf 99}, no. 8, 085005 (2019)
  arXiv:1806.02835 [hep-th].

  \bibitem{Gapp}
  J.~S.~Cotler, J.~Molina-Vilaplana and M.~T.~Mueller,
  arXiv:1612.02427 [hep-th].
  \bibitem{Gapp1}
T.~Barnes and G.~I.~Ghandour,
Phys.\ Rev.\ D {\bf 22}, 924 (1980).

\bibitem{Gapp2}
  P.~M.~Stevenson,
  Phys.\ Rev.\ D {\bf 32}, 1389 (1985).


\bibitem{Gapp3}
P.~M.~Stevenson and I.~Roditi,
  Phys.\ Rev.\ D {\bf 33}, 2305 (1986).
\bibitem{Gapp4}
W.~Namgung, P.~M.~Stevenson and J.~F.~Reed,
  Z.\ Phys.\ C {\bf 45}, 47 (1989).
\bibitem{Gapp5}
F.~Siringo and L.~Marotta,
  Phys.\ Rev.\ D {\bf 78}, 016003 (2008)
  [arXiv:0803.3043 [hep-ph]].
  


  \bibitem{AB} 
  A.~Bhattacharyya, A.~Shekar and A.~Sinha,
  JHEP {\bf 1810}, 140 (2018),
  arXiv:1808.03105 [hep-th].
  
\bibitem{JM13}
 V.~Balasubramanian, M.~DeCross, A.~Kar and O.~Parrikar,
  JHEP {\bf 1902} (2019) 069, [arXiv:1811.04085 [hep-th]].

 
  
  
  








 
 



\end{thebibliography}
\end{document}